\documentclass[aps,prl,reprint,floatfix,notitlepage, superscriptaddress]{revtex4-2}
\usepackage[utf8]{inputenc}
\usepackage{graphicx}
\usepackage{amsmath}
\usepackage{amssymb}
\usepackage{commath}
\usepackage{braket}
\usepackage{hyperref}
\usepackage{subfigure}
\usepackage[usenames,dvipsnames]{color}

\usepackage{booktabs} 
\usepackage{tabularx} 
\usepackage{array} 
\usepackage{longtable}

\usepackage[mathscr]{euscript}
\usepackage{bm}
\usepackage{dcolumn}
\usepackage{tikz,pgfplots}
\pgfplotsset{compat=1.18}
\usetikzlibrary{decorations.pathmorphing,patterns,positioning}
\usetikzlibrary{tikzmark}
\usetikzlibrary{shapes,shadows,arrows}

\usepackage{xcolor}
\definecolor{gatecolor}{HTML}{6FA4FF}

\newcommand\beq{\begin{equation}}
\newcommand\eeq{\end{equation}}
\newcommand\bea{\begin{eqnarray}}
\newcommand\eea{\end{eqnarray}}

\newcommand{\Tr}{{\rm Tr}}

\newcommand{\identity}{\mathbb{\hat{I}}}

\usepackage{comment}
\usepackage[normalem]{ulem}

\begin{document}
	
\title{Clifford Ergotropy}

\author{ Somnath Maity}
\affiliation{Nonequilibrium Quantum Statistical Mechanics RIKEN Hakubi Research Team, RIKEN Pioneering Research Institute (PRI), 2-1 Hirosawa, Wako, Saitama 351-0198, Japan}

\author{Ryusuke Hamazaki}
\affiliation{Nonequilibrium Quantum Statistical Mechanics RIKEN Hakubi Research Team, RIKEN Pioneering Research Institute (PRI), 2-1 Hirosawa, Wako, Saitama 351-0198, Japan} 
\affiliation{RIKEN Center for Interdisciplinary Theoretical and Mathematical Sciences (iTHEMS), 2-1 Hirosawa, Wako, Saitama 351-0198, Japan}
    
\date{\today}

\begin{abstract}
We discuss the interplay between thermodynamics and magic resources in closed quantum dynamics by introducing Clifford ergotropy, the amount of extractable energy under the restriction to Clifford operations. We provide universal upper bounds on Clifford ergotropy, which decrease with increasing magic as quantified by the infinite-order filtered stabilizer Rényi entropy.
We demonstrate the utility of this bound for one- and two-qubit systems, with the latter exhibiting a notable transition in the control landscape of Clifford ergotropy.
Finally, we show that our analysis has nontrivial consequences even for many-body systems where the exact optimization is generally difficult to perform, including a form of the second law of thermodynamics under Clifford operations for typical quantum states.
\end{abstract}

\maketitle

{\it Introduction--}
Recent developments in techniques for precisely controlling quantum systems have motivated the study of thermodynamics in the quantum regime~\cite{vinjanampathy2016quantum,goold2016role,binder2018thermodynamics,Campbell_2026}.
In particular, the extraction and charging of energy under quantum operations have attracted extensive recent interest~\cite{Oppenheim2002,Skrzypczyk2014,Perarnau-Llobet2015,Andolina2019,watanabe2026}, not only  fundamentally but also in relation to new quantum devices known as quantum batteries~\cite{Campaioli2024}.
From a fundamental viewpoint, passive states~\cite{lenard1978thermodynamical,pusz1978passive} are defined as states from which no work can be extracted by any unitary evolution, thereby providing a form of the second law of thermodynamics in closed quantum systems.
In contrast, the maximum work extractable from a non-passive state is known as ergotropy~\cite{Allahverdyan_2004}, and its properties have been actively studied for decades.
In particular, while ergotropy is originally defined for arbitrary unitary operations, it has become an interesting topic how the extractable work changes under restricted control~\cite{Perarnau-Llobet2015,brown2016passivity,Francica20,Garcia20,Shi2022,Mitsuhashi2022,Puliyil2022,rodriguez2025extracting,sugimoto2025optimal,Hokkyo2025,Polo26}.

Despite these developments, little is known about how restricted extractable work is constrained by quantum magic, a fundamental resource that has gathered considerable recent attention.
Quantum magic, or nonstabilizerness, quantifies the amount of non-Clifford resources of quantum states distinct from the stabilizer framework~\cite{Bravyi2005,Veitch2014}. According to the Gottesman–Knill theorem~\cite{Gottesman1998,Aaronson2004}, stabilizer states generated by Clifford operations can be efficiently simulated on classical computers, despite exhibiting a high degree of entanglement. 
Non-Clifford resources therefore offer a crucial resource for universal quantum computation beyond entanglement. 
Among several possible magic monotones~\cite{Bravyi2016rank,Bravyi2019fidelity,Howard2017rom},
stabilizer Rényi entropies (SRE) have recently emerged as a scalable measure of magic, particularly for pure states~\cite{leone2022,leone2024}. This has enabled extensive studies of magic in many-body settings, including quantum phases and phase transitions~\cite{Huag2023quantifying,Tarabunga2023,Hoshino2026,Hoshino2026sretopology}, nonequilibrium dynamics~\cite{Turkeshi2025magicspreading,turkeshi2025,Magni2025,Odaviifmmode2025,Tarabunga2025magic,Maity2026,Bejan2025magicspreading}, and experimental protocols~\cite{Oliviero2022,Haug2023,Haug2024}.

In this Letter, we introduce Clifford ergotropy, the maximum work extractable under Clifford operations alone, and reveal novel interplay between quantum thermodynamics and magic resources.
We first provide general upper bounds on Clifford ergotropy through the permutation structure of the Pauli coefficients.
Interestingly, {if we compare pure states with the same initial energies}, the bounds decrease as the infinite-order filtered SRE increases, indicating that magic in the state can hinder energy extraction under Clifford operations.
We then demonstrate the utility of our bound for one- and two-qubit systems.
For a particular one-qubit case, we show that Clifford ergotropy can be expressed in terms of the stabilizer fidelity; for the two-qubit case, it exhibits a notable  transition in the control landscape, which is also captured by our bound.
Finally, we discuss nontrivial consequences of our analysis for many-body systems, where the exact optimization is generally difficult to perform.
In particular, we demonstrate that our bound can become useful for multi-qubit product states.
Furthermore, we prove that Clifford ergotropy macroscopically vanishes for typical quantum states owing to their large amount of magic, thereby providing a form of the second law of thermodynamics in closed quantum many-body systems under Clifford-restricted operations.

{\it Clifford ergotropy--} 
The ergotropy is the maximum amount of work that can be extracted from a nonpassive state using a cyclic unitary operation~\cite{Allahverdyan_2004}.  
Given an initial (and final) Hamiltonian $\hat{H}$ and a density matrix $\hat{\rho}$ in the Hilbert space whose dimension $d$ is finite, the ergotropy is defined as 
\begin{eqnarray}
    \mathcal{E}(\hat{\rho}) = E(\hat{\rho}) - \min_{U}\left\{ E (U \hat{\rho} U^\dag)\right\},
\end{eqnarray}
where $E(\hat{\varrho})=\Tr[\hat{\varrho}\hat{H}]$ denotes the energy expectation value of the system in state $\hat{\varrho}$ and the minimization is performed over all possible unitaries $U$. Equivalently,
$\mathcal{E}(\hat\rho)=\Tr\left[ \hat{H}(\hat{\rho}- \hat{\pi}_\rho) \right]$, where $\hat{\pi}_\rho$ is called the passive state, from which no further work can be obtained via unitary operations. If we decompose the density matrix and the Hamiltonian as $\hat{\rho} = \sum_{k=1}^{d} p_k \ket{p_k}\bra{p_k}$ with $p_k \geq p_{k+1}$ and $\hat{H} = \sum_{i=1}^{d} \varepsilon_i \ket{\varepsilon_i}\bra{\varepsilon_i}$ with $\varepsilon_i \leq \varepsilon_{i+1}$, respectively, the passive state is represented as
$\hat{\pi}_\rho = \sum_{k=1}^d p_k \ket{\varepsilon_k}\bra{\varepsilon_k}$. 
That is, we perform the permutation of the occupation probabilities such that the larger probability corresponds to lower energies.
Then, the ergotropy is given as $\mathcal{E}(\hat{\rho}) = E(\hat{\rho}) -\sum_i p_i \varepsilon_i$.
In particular, it becomes $\mathcal{E}(\hat{\rho}) = E(\hat{\rho}) -\varepsilon_G$ for pure states, where $\varepsilon_G=\varepsilon_1$ is the ground-state energy of $\hat{H}$.

We now introduce a resource-theoretic refinement of ergotropy in which the extracting unitary is restricted to the Clifford operations $C$ only, instead of all possible unitary operations. The ergotropy restricted to those Clifford operations is defined as
\begin{eqnarray}
    \mathcal{E}_\mathrm{Cl}(\hat{\rho}) = E(\hat{\rho}) - \min_{C}\left\{ E\left( C \hat{\rho} C^\dag\right)\right\},
\end{eqnarray}
which we call as \textit{Clifford ergotropy}. The quantity $\min_{C}\{ E\left( C \hat{\rho} C^\dag\right)\}$ is the minimum energy on the Clifford orbit of $\hat{\rho}$. 

For an arbitary state $\hat{\rho}$, $\mathcal{E}_\mathrm{Cl}(\hat{\rho}) \leq \mathcal{E}(\hat{\rho})$,
which naturally motivates us to define an ergotropy gap as the difference between unrestricted and Clifford work extraction:
\begin{eqnarray}
    \Delta\mathcal{E} = \mathcal{E}(\hat{\rho}) - \mathcal{E}_\mathrm{Cl}(\hat{\rho})\geq 0.
\end{eqnarray}
In particular, for an initial pure state, we find
\begin{eqnarray}
    \Delta\mathcal{E} = -\varepsilon_G +\min_{C}\left\{ E\left( C \hat{\rho} C^\dag\right)\right\}.
\end{eqnarray}
Operationally, $\Delta\mathcal{E}$ captures the amount of work unlocked in the presence of the non-Clifford  {operations}.
As an example, $\Delta\mathcal{E}=0$ if $\hat{\rho}$ is a stabilizer pure state, and the ground state of $\hat{H}$ is also a stabilizer state.
In contrast, if $\hat{\rho}$ is a stabilizer (non-stabilizer) pure state, and the ground state of $\hat{H}$ is a non-stabilizer (stabilizer) state, we have $\Delta \mathcal{E}>0$.

{Interestingly, if $\hat{\rho}$ is a stabilizer pure state, the gap $\Delta\mathcal{E}$ coincides with the ``stabilizer gap" introduced in Ref.~\cite{macedo2026every}, which
measures how much the lowest-energy sector of $\hat H$ is inaccessible to the
stabilizer states; however, if $\hat{\rho}$ is a non-stabilizer state, these two gaps are conceptually distinct.
We also note} that $\Delta\mathcal{E}$ differs from the ergotropy gap defined from the difference between ergotropy under global and local unitaries, which has been studied as a thermodynamic signature of quantum correlations~\cite{Mukherjee2016,Alimuddin2019,Puliyil2022,Polo26}.

In the following, we focus on a system of $N$ qubits.
Since the $N$-qubit Pauli strings form an orthogonal operator basis on the Hilbert space $\mathcal{H}$ of dimension $d=2^N$, both $\hat{H}$ and $\hat{\rho}$ can be expanded in this basis. Let $\mathcal{P}_N = \{\hat{P}_{\mu}\}_{\mu=0}^{d^2-1}$ denote the set of Pauli strings, with $\hat{P}_0=\identity$
and $\hat{P}_\mu = \hat\sigma_{0}\otimes\hat\sigma_{1}\otimes\cdots\otimes\hat\sigma_{N-1}$, where $\hat\sigma_j = \{\identity_j, \hat{X}_j, \hat{Y}_j, \hat{Z}_j\}$. The basis satisfies $\Tr[\hat{P}_\mu \hat{P}_\nu]=d\delta_{\mu\nu}$.
First, the state can be expanded as $\hat{\rho} = (1/d)\sum_{\mu=0}^{d^2-1} \rho_\mu \hat{P}_\mu$, with coefficients $\rho_\mu=\Tr[\hat{\rho} \hat{P}_\mu]$. Here, $\{\rho_\mu\}_{\mu=0}^{d^2 -1}$ is called the Pauli coefficients, which constitute the Pauli spectrum $\{|\rho_\mu|^2/d\}$~\cite{leone2022,schuster2024paulispectrum} of the state $\rho$.
Next, the Hamiltonian is assumed to be traceless without loss of generality,
since the constant term does not affect the (standard or Clifford) ergotropy.
We can then write $\hat{H} = \sum_{\mu=1}^{K}H_{\mu}  \hat{P}_\mu $ with coefficients $H_\mu =\Tr[\hat{H} \hat{P}_\mu]/d$, where $K$ is the number of nonzero coefficients.
Importantly, $K$ is much smaller than $d^2-1$ for many-body systems with few-body interactions.
For example, $K=\mathrm{O}(N)$ for short-range Hamiltonians.

Clifford conjugation maps Pauli strings to signed Pauli strings, 
\begin{eqnarray}
    C^\dag \hat{P}_\mu C = \eta_{\mu}^C \hat{P}_{c(\mu)},  
\end{eqnarray}
where $\eta_{\mu}^C=\pm1$ and the permutated label $c(\mu)$ depend on $C$, and it preserves all the commutation and anticommutation relations between the Pauli strings. In terms of the Pauli coefficients of the density matrix and the Hamiltonian coefficients, we have
\begin{eqnarray}
    E(C \hat \rho C^\dag) &=&  \sum_{\ell=1}^{K} \eta_\ell^C H_\ell  \rho_{c(\ell)}.
\end{eqnarray}

From the above, the Clifford ergotropy is obtained by  the optimization by a constrained permutation of the Pauli coefficients.
Note that, while the permutation of the Pauli coefficients makes interesting analogy to that of the spectrum $p_k$ of $\hat\rho$ in obtaining the standard ergotropy, the former permutation is not arbitrary, which generally makes exact calculation of the Clifford ergotropy nontrivial. Still, we can find simple upper bounds by relaxing the constrained permutation among the Pauli coefficients to the arbitrary one, as discussed below.

{\it Bound on Clifford ergotropy--} 
To find upper bounds on $\mathcal{E}_\mathrm{Cl}(\rho)$, let \(r_1\ge r_2\ge\cdots\ge r_{d^2-1}\) be the nonincreasing rearrangement of
\(\{|\rho_\mu|: \hat{P}_\mu\in\mathcal P_N\setminus\{\mathbb I\}\}\), and let
\(h_1\ge h_2\ge\cdots\ge h_K>0 =h_{K+1}=\cdots=h_{d^2-1}\) be the nonincreasing rearrangement of
\(\{|H_\ell|\}_{\ell=1}^K\) with additional $d^2-K-1$ zeros.
Thus, \(r_\mu\) and \(h_\mu\) are, respectively, the \(\mu\)th largest absolute Pauli coefficient of the state and the \(\mu\)th largest absolute Pauli coefficient of the Hamiltonian. If one ignored the commutation-preservation constraint and allowed an arbitrary permutation, the corresponding minimum energy would be given by $-\mathbf{r}\cdot\mathbf{h}=-\sum_{\ell=1}^{K}r_\ell h_\ell $, where $\mathbf{v}=(v_1,\cdots,v_{d^2-1})$. 
Consequently, the actual minium energy should be constrained as
\begin{align}
    \min_{C \in \mathcal{C}_N}\left\{ E\left( C \hat{\rho} C^\dag\right)\right\} \ge -\mathbf{r}\cdot\mathbf{h}=-\sum_{\ell=1}^{K}r_\ell h_\ell.
\end{align}

Therefore, we have the  upper bound on $\mathcal{E}_\mathrm{Cl}(\hat\rho)$ as our main result:
\begin{eqnarray}\label{eq:bound_coeff}
    \mathcal{E}_\mathrm{Cl}(\hat{\rho}) \le E(\hat{\rho}) +\mathbf{r}\cdot\mathbf{h}.
\end{eqnarray}
Furthermore, using the H\"{o}lder's inequality, we have
\begin{eqnarray}\label{eq:holder}
\mathcal{E}_\mathrm{Cl}(\hat{\rho}) \le E(\hat{\rho}) + r_1 \|\mathbf H\|_1
\end{eqnarray}
and,  for pure states,
\begin{eqnarray}
\Delta \mathcal{E}\ge -\varepsilon_G - \mathbf{r}\cdot\mathbf{h}\ge -\varepsilon_G - r_1 \|\mathbf H\|_1,
\end{eqnarray}
where 
\begin{align}
\|\mathbf H\|_1 = \sum_\ell |H_\ell|=\sum_\ell h_\ell
\end{align}
is the $L_1$-norm of the Hamiltonian.

Next,  {focusing on pure states}, let us describe our bound by one of the magic measures, i.e., the infinite-order filtered SRE.
Using the Pauli spectrum 
$\{r_\mu^2/d\}_{\mu=0}^{d^2-1}$, we can define a filtered SRE~\cite{leone2022,turkeshi2025} as
$
    {M}_\alpha = \frac{1}{1-\alpha} \ln\left( \sum_{\mu=1}^{d^2-1} \frac{r_\mu^{2\alpha}}{d-1} \right).
$
Here, the contribution from the identity is removed and the filtered spectrum is appropriately normalized as $\sum_{\mu=1}^{d^2-1} r_\mu^2/(d-1)=1$ such that ${M}_\alpha=0$ for the stabilizer states~\cite{turkeshi2025}. 
Then, Eq.~\eqref{eq:holder} is written as 
\begin{eqnarray}\label{eq:bound_fsre}
    \mathcal{E}_\mathrm{Cl}(\hat{\rho}) \le E(\hat{\rho}) + e^{-{M}_\infty /2} \|\mathbf{H}\|_1,
\end{eqnarray}
where 
\begin{align}
{M}_\infty = -\ln[r_1^2]
\end{align}
is the infinite-order filtered SRE ($\alpha\rightarrow\infty$).

Remarkably, the inequality~\eqref{eq:bound_fsre}, stating that the upper bound becomes smaller if the magic becomes larger   {(for a fixed initial energy $E(\hat\rho)$)},
indicates that the magic can be an obstacle for work extraction.
In general, higher-magic states do not necessarily have smaller Clifford ergotropy than lower-magic states because the magic appears as the bound rather than equality.
However, as discussed below, we indeed find many examples that demonstrate the decrease of the Clifford ergotropy for high magic states.

{\it Single-qubit case--} 
We first illustrate the Clifford ergotropy in the simplest case of a single qubit where we obtain the equality between $\mathcal{E}_\mathrm{Cl}$ and magic, as a special case of the general bounds. We especially focus on  $\hat{H}=h \hat Z$ and expand the state as $\hat{\rho}=\frac{1}{2}\left(\mathbb I+\rho_x\hat{X}+\rho_y\hat{Y}+\rho_z\hat{Z}\right)$.
Since the Clifford orbit of the Hamiltonian contains only the signed Pauli operators $ C\hat ZC^\dag \in \{ \pm \hat X, \pm \hat Y, \pm \hat Z \}$,
the minimum energy reachable by the Clifford operations is 
$
    h\min_C \Tr[\hat ZC\hat{\rho}C^\dag] = -h r_1,
$
where $ r_1 = \max \{\abs{\rho_x},\abs{\rho_y},\abs{\rho_z} \}$. It follows that $\mathcal{E}_\mathrm{Cl}(\hat{\rho}) = h(\rho_z + r_1)$ and $\mathcal{E}(\hat{\rho}) = h(\rho_z + |\boldsymbol{\rho}|)$,
where $\boldsymbol{\rho}=(\rho_x,\rho_y,\rho_z)$ is the Bloch vector.
The ergotropy gap is therefore given by $\Delta\mathcal{E} = h(|\boldsymbol{\rho}| - r_1)$.
This single-qubit result offers a special case of our general bounds in \eqref{eq:bound_coeff} and \eqref{eq:holder} for which the equality condition is achieved. Indeed, we have $\|\mathbf{H}\|_1=h$ in this case.

Notably, the single-qubit ergotropy gap is also directly related to the single-qubit stabilizer fidelity~\cite{Bravyi2019fidelity}
\begin{eqnarray}
    F_{\rm STAB}(\hat{\rho}):= \max_{|s\rangle\in{\rm STAB}}\langle s|\hat{\rho}|s\rangle = \frac{1}{2}\left(1+r_1\right).
\end{eqnarray}
For a single-qubit pure state $|\boldsymbol{\rho}|=1$, one obtains
\begin{equation}
    \Delta\mathcal{E} = 2h\left(1- F_{\rm STAB}(\hat{\rho})\right)=2h\left(1-e^{-D_{\min}(\hat{\rho}\Vert{\rm STAB})}\right),
\end{equation}
where $D_{\min}(\hat{\rho}\Vert{\rm STAB}):=-\ln F_{\rm STAB}(\hat{\rho})$ is a magic monotone known as \textit{min-relative entropy of magic}. For a pure stabilizer state $F_{\mathrm{STAB}}(\rho)=1$ and $\Delta\mathcal{E}=0$, and the ergotropy gap is a direct witness of magic. Note that for a mixed state, the gap instead reads $\Delta \mathcal{E}=h(1+|\boldsymbol{\rho}|-2F_{\rm STAB}(\hat{\rho}))$ and is no longer a faithful witness of magic. 

{\it Two-qubit case and the Clifford ergotropy transition--} The discrete nature of the Clifford optimization can produce sharp changes in the optimal work-extraction protocol, resulting in a notable transition of Clifford ergotropy. We illustrate this using a simple two-qubit transverse field Ising Hamiltonian 
\begin{align}\label{eq:twoHam}
\hat{H}= -\hat{Z}_1\hat{Z}_{2}+g(\hat{X}_1+\hat{X}_2)+h(\hat{Z}_1+\hat{Z}_2). 
\end{align}

\begin{figure}[]
    \centering
    \includegraphics[width=0.45\textwidth,height=.3\textwidth]{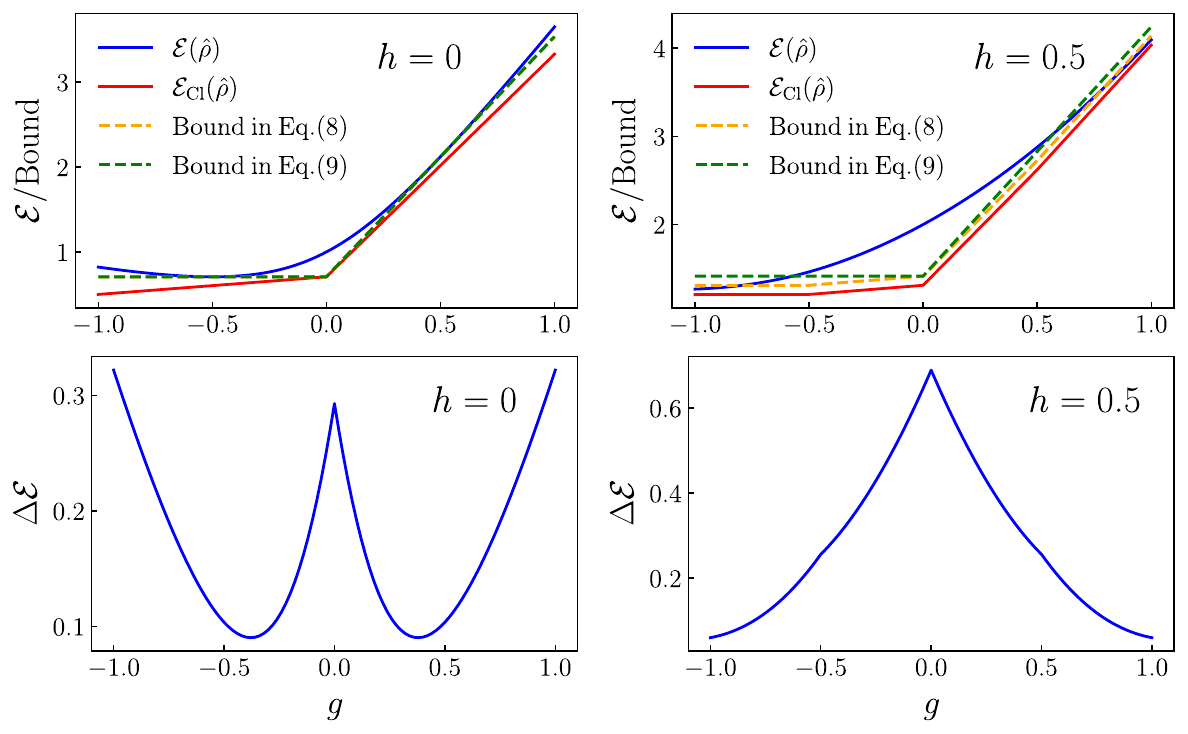}
    \caption{ Comparison between the ergotropy $\mathcal{E}(\hat{\rho})$, the Clifford ergotropy $\mathcal{E}_\mathrm{Cl}(\hat{\rho})$, and the analytical bounds as a function of the transverse field $g$ for the two-qubit Hamiltonian in Eq.~\eqref{eq:twoHam}. The upper panels show $\mathcal{E}(\hat{\rho})$, $\mathcal{E}_{\rm Cl}(\hat{\rho})$, and the bounds in Eqs.~\eqref{eq:bound_coeff} and \eqref{eq:holder} for $h=0$ and $h=0.5$, where the two bounds coincide for $h=0$. The lower panels show the corresponding gap $\Delta\mathcal{E} = \mathcal{E}(\hat{\rho}) - \mathcal{E}_\mathrm{Cl}(\hat{\rho})$. The Clifford ergotropy and the ergotropy gap exhibit transitions in the control landscape at $g=0$ for $h=0$, and at $g=0$ and $|g|=0.5$ for $h=0.5$, which are also captured by the bound in Eq.~\eqref{eq:bound_coeff}. The initial state is $\hat\rho=\ket{TT}\bra{TT}$, where $\ket{T}$ is the T state.
    }
    \label{fig:ergotropy_comparison}
\end{figure}

The Clifford ergotropy $\mathcal{E}_{\rm Cl}(\hat{\rho})$ for this Hamiltonian 
with the state $\hat\rho=\ket{TT}\bra{TT}$, where $\ket{T}=\frac{1}{\sqrt{2}}(\ket{0}+e^{i\pi/4}\ket{1})$ is the T state,
is shown in Fig.~\ref{fig:ergotropy_comparison} as a function of $g$ {[see End matter for the analytical expression]}, together with the full ergotropy $\mathcal{E}(\hat{\rho})$. We also plot the bounds derived in Eqs.~\eqref{eq:bound_coeff} and \eqref{eq:holder}. For zero longitudinal field $h=0$, the two bounds coincide, whereas for a nonzero longitudinal field $h=0.5$, the bound in Eq.~\eqref{eq:bound_coeff} is tighter than that in Eq.~\eqref{eq:holder}. The bounds become nontrivial and even achieve the equality condition (e.g., at $h=g=0$), while they can be trivial for some parameter regime (e.g., near $\abs{g}=1$ for $h=0.5$). The lower panels of Fig.~\ref{fig:ergotropy_comparison} show the ergotropy gap $\Delta\mathcal{E}$. 

Interestingly, the Clifford ergotropy and the gap exhibit transitions in the control landscape~\cite{Beato25L,Beato25X} at $g=0$, $h=0$ and $\abs{g} + \abs{h}=1$ [see the End Matter],
represented as their cusps. Here, such transitions of the control landscape occur because the optimal Clifford operator achieving $\mathcal{E}_\mathrm{Cl}$ changes abruptly among the finite number of elements due to the change of the parameters.
This is in contrast with the standard ergotropy, where no such transitions appear there. Notably, the tighter bound in Eq.~\eqref{eq:bound_coeff} correctly captures these transitions, showing the small cusps at $g=0,\pm 0.5$.

{\it Many-body case and the second law--}
Our analysis is useful even for many-body systems, where the exact optimization is generally difficult.
Let us first consider product states given in the form $\hat\rho_\mathrm{prod}=\bigotimes_{j=1}^N\hat\rho'_j$, where $\hat\rho_j'$ are single-site density matrices.
If we denote the absolute value of the ordered Pauli coefficients as $1,r_{j1},r_{j2}\cdots$, we find $r_1=\max_{j}r_{j1}$. Therefore, we have
\begin{align}
\mathcal{E}_\mathrm{Cl}(\hat{\rho}_\mathrm{prod}) \leq E(\hat{\rho}_\mathrm{prod})+ \mathbf{r}\cdot \mathbf{h}\leq E(\hat{\rho}_\mathrm{prod})+\left(\max_{j}r_{j1}\right)\|\mathbf{H}\|_1
\end{align}
and, for pure states, the ergotropy gap is bounded as
\begin{align}
\Delta \mathcal{E}\geq -\varepsilon_G-\mathbf{r}\cdot \mathbf{h}\geq  -\varepsilon_G-\left(\max_{j}r_{j1}\right)\|\mathbf{H}\|_1.
\end{align}

For example, if we consider a tensor product of $T$-states, $\ket{\psi_\mathrm{prod}}=\ket{T\cdots T}$, we find $\max_jr_{j1}=1/\sqrt{2}$.
For the case with the one-dimensional  classical Ising model under the periodic boundary condition, $\hat{H}=-\sum_{j=1}^N\hat{Z}_j\hat{Z}_{j+1}+h\hat{Z}_j$, we have $\Delta \mathcal{E}\geq N\left(1-\frac{1}{\sqrt{2}}\right)(1+|h|)>0$, which always gives a positive non-trivial bound.
For the case with the transverse-field Ising model $\hat{H}=-\sum_{j=1}^N\hat{Z}_j\hat{Z}_{j+1}+g\hat{X}_j$~\cite{pfeuty1970one} for large $N$, we have 
$\Delta \mathcal{E}\gtrsim N(1+|g|)\left(\frac{2}{\pi}\mathfrak{E}\left(\frac{2\sqrt{|g|}}{1+|g|}\right)-\frac{1}{\sqrt{2}}\right)$ with $\mathfrak{E}(x)=\int_0^{\pi/2}dk\sqrt{1-x^2\sin^2 k}$ being the complete elliptic integral of the second kind.
This provides a positive nontrivial bound for $|g|\lesssim 0.506$ and $|g|\gtrsim 1.975$.

Next, let us consider pure states with an extensive infinite-order filtered SRE, $M_\infty=\mathrm{O}(N)$ (or $r_1=e^{-\mathrm{O}(N)}$). 
For Hamiltonians with short-range interactions, 
our bound in  \eqref{eq:bound_fsre} leads to the conclusion that the Clifford ergotropy is exponentially small while the full ergotropy is extensively large for such states, 
\begin{align}
\mathcal{E}_\mathrm{Cl}(\hat{\rho}_\mathrm{typ})=e^{-\mathrm{O}(N)}, \quad \Delta \mathcal{E}=\mathrm{O}(N).
\end{align}
Here, we have used that the first term in the bound in  \eqref{eq:bound_fsre} is also surpressed as $|E(\hat\rho)|\leq \left|\sum_{\mu}\rho_\mu H_\mu\right| \leq r_1\|\mathbf{H}\|_1=e^{-\mathrm{O}(N)}$. 
Our result indicates a form of the second law of thermodynamics of closed quantum many-body systems from the infinite-temperature pure states, i.e., no work can be extracted from Clifford operations alone for sufficiently high-magic states.

One of the classes of states satisfying the above is typical pure states $\hat{\rho}_\mathrm{typ}=\ket{\psi_\mathrm{typ}}\bra{\psi_\mathrm{typ}}$ chosen according to the unitary Haar measure.
In fact, we can show that the maximum Pauli coefficient $r_1$ is exponentially small in this case (indeed, we find $M_\infty\simeq N\ln 2$ for large $N$), as shown in End matter.

Furthermore, we can also prove another form of the second law at finite energy density (temperature) for normal macroscopic systems: an extensive energy cannot be extracted via Clifford operations from a typical state $\hat{\rho}_\mathrm{typ}^{E_0}=\ket{\psi_\mathrm{typ}^{E_0}}\bra{\psi_\mathrm{typ}^{E_0}}$ drawn from the microcanonical energy shell at energy ${E_0}\:(\leq 0)$.
Instead of using the bound, we prove 
\begin{align}
\mathcal{E}_\mathrm{Cl}(\hat{\rho}_\mathrm{typ}^{E_0})=\mathrm{o}(N), \quad \Delta \mathcal{E}=\mathrm{O}(N)
\end{align}
from the following facts.
First, the Pauli coefficients of a typical state and those of microcanonical ensemble $\hat{\rho}_\mathrm{mic}^{E_0}$ are exponentially close, leading to $E(C\hat{\rho}_\mathrm{typ}^{E_0}C^\dag)\simeq E(C\hat{\rho}_\mathrm{mic}^{E_0}C^\dag)$ for any Clifford operator $C$ (see End matter). Taking $C_m$ that minimizes $E(C\hat{\rho}_\mathrm{typ}^{E_0}C^\dag)$ and using $E(\hat{\rho}_\mathrm{typ}^{E_0})\simeq E(\hat{\rho}_\mathrm{mic}^{E_0})$, we have
$\mathcal{E}_\mathrm{Cl}(\hat{\rho}_\mathrm{typ}^{E_0})\simeq E(\hat{\rho}_\mathrm{mic}^{E_0})-E(C_m\hat{\rho}_\mathrm{mic}^{E_0}C_m^\dag)\leq \mathcal{E}_\mathrm{Cl}(\hat{\rho}_\mathrm{mic}^E)$.
Finally, since an extensive energy cannot be extracted from the microcanonical ensemble for normal macroscopic systems, i.e., $\mathcal{E}_\mathrm{Cl}(\hat{\rho}_\mathrm{mic}^E)\leq \mathcal{E}(\hat{\rho}_\mathrm{mic}^E)=\mathrm{o}(N)$~\cite{tasaki2000statistical}, we have $\mathcal{E}_\mathrm{Cl}(\hat{\rho}_\mathrm{typ}^{E_0})=\mathrm{o}(N)$.

We remark that our second law in closed quantum many-body systems relies on that the typical state has sufficiently high magic and do not impose any assumptions for the operational times as previously considered in Refs.~\cite{goldstein2013second,ikeda2015second,Kaneko2017,Hokkyo2025,chiba2026second}.
We also note that the second law under Clifford operations may not apply for locally thermal but non-typical states, such as the entangled antipodal
pair (EAP) states~\cite{chiba2024} corresponding to the infinite temperature. In fact, EAP states are stabilizer states, and if the Hamiltonian has a stabilizer ground state, we can extract full ergotropy only through Clifford operations.

{\it Conclusion--}
We have connected  {ergotropy} and magic resource for the first time by introducing a notion of Clifford ergotropy.
The universal bounds on Clifford ergotropy allow us to analyze how the magic of the initial states affects the extractable work under the restriction, as demonstrated in one, two, and even many-qubit systems.
In addition, the Clifford ergotropy exhibits notable features such as the transitions in the control landscape and the second law of thermodynamics in closed quantum many-body systems.
Note that our bounds can also be applied to evaluate minimizing (or maximizing) an average of arbitrary observables $O$, other than $H$.
In particular, by setting $O=-H$, our bounds become useful to understand the maximum energy that can be charged in a quantum battery under Clifford restrictions.

Our work opens interesting future directions. 
While we have demonstrated that the bounds often become useful, it is interesting to see if we can find a better bound or an efficient optimization algorithm to obtain the Clifford ergotropy.
Another important direction is to generalize our framework to non-unitary dynamics, which involves, e.g., Clifford measurement {or thermal baths~\cite{hwfq-ytkb,macedo2026every}}.
Finally, it is intriguing to elucidate if our analysis extends to general quantum resource theories~\cite{Chitamber2019,Takagi2019,Lostaglio_2019} other than magic.

{\it Note added--} After completion of our work, we became aware of Ref.~\cite{konar2026}, which also discusses the relation between ergotropy and magic from a different perspective.

\begin{acknowledgments}
{\it Acknowledgments--} 
    The authors are grateful to Kohei Yoshimura and Yuuya Chiba for fruitful discussions. We also thank  { Alexssandre De Oliveira Junior} for helpful comments on the manuscript. {We are thankful to Shoki Sugimoto for letting us know the analytical expression for the two-qubit case.}
    R.H.\ is supported by JSPS KAKENHI Grant No.~JP24K16982.
    S.M.\ and R.H.\ are supported by JST ERATO Grant No.~JPMJER2302, Japan.
\end{acknowledgments}

\textit{Data availability--}
All the data that support the plots in this manuscript are available from the authors upon reasonable request.

\bibliography{references_ergotropy.bib}

\section{End Matter}

\section{Analytical expression for the two-qubit case}
Here, we derive the analytical expression for the Clifford ergotropy of the two-qubit Hamiltonian in Eq.~\eqref{eq:twoHam}. For the initial state $\hat\rho=\ket{TT}\bra{TT}$,  the nonzero Pauli coefficients are 
\begin{eqnarray}
    \rho_{\hat{X}_1} = \rho_{\hat{Y}_1} = \rho_{\hat{X}_2} = \rho_{\hat{Y}_2} = \frac{1}{\sqrt{2}} \equiv a, \\
    \rho_{\hat{X}_1\hat{X}_2} = \rho_{\hat{X}_1\hat{Y}_2} = \rho_{\hat{Y}_1\hat{X}_2} = \rho_{\hat{Y}_1\hat{Y}_2} = \frac{1}{2} \equiv b.
\end{eqnarray}
The rest of the Pauli coefficients corresponding to Pauli strings containing at least one $\hat Z$ are zero. For notational simplicity, we explicitly mention the two-qubit Pauli strings instead of the integer index $\mu$ unlike the main text. Therefore, we have 
\begin{eqnarray}\label{eq:r_tt}
    \mathbf{r}=(a,a,a,a,b,b,b,b,0,0,0,0,0,0,0).
\end{eqnarray}

Let
\begin{eqnarray}
    \hat A_j=C^\dagger \hat X_j C,\qquad
\hat B_j=C^\dagger \hat Z_j C
\end{eqnarray}
be the Clifford transformations of the Pauli generators. Then
\begin{eqnarray}
    C^\dagger \hat Z_1\hat Z_2 C=\hat B_1\hat B_2,
\end{eqnarray}
up to the corresponding Clifford sign. Since Clifford operations preserve
the Pauli commutation relations, the operators $\hat A_1,\hat B_1,\hat A_2,\hat B_2$ satisfy
\begin{eqnarray}
    \{\hat A_j,\hat B_j\}=0,
\qquad
[\hat A_j,\hat B_k]=0\quad (j\neq k).
\end{eqnarray}

Note that the overall sign for both the transverse-field sector $g( \hat X_1 + \hat X_2)$ and the longitudinal-field sector $h (\hat Z_1 + \hat Z_2)$ can be chosen independently from other terms. Indeed,
\begin{eqnarray}
    \hat Z_1\hat Z_2(\hat X_1+\hat X_2)\hat Z_1\hat Z_2 = -(\hat X_1+\hat X_2),
\end{eqnarray}
while $\hat Z_1\hat Z_2$ leaves $\hat Z_1+\hat Z_2$ and
$\hat Z_1\hat Z_2$ invariant. Similarly,
\begin{eqnarray}
    \hat X_1\hat X_2(\hat Z_1+\hat Z_2)\hat X_1\hat X_2 = -(\hat Z_1+\hat Z_2),
\end{eqnarray}
while $\hat X_1\hat X_2$ leaves $\hat X_1+\hat X_2$ and
$\hat Z_1\hat Z_2$ invariant. Hence, the minimum energy on the Clifford orbit depends on $\abs{g}$ and $\abs{h}$. Therefore, the Clifford optimization task is equivalent to assigning the components of the vector $\mathbf{r}$ in Eq.~\eqref{eq:r_tt} to the Hamiltonian terms $(\hat Z_1 \hat Z_2, \hat X_1, \hat X_2, \hat Z_1, \hat Z_2)$ optimally.

Now, in the following, we show that the Clifford optimization has only two branches. First, suppose we asign the Pauli coefficient $b$ to the interaction term,
\begin{eqnarray}
    \Tr [\hat Z_1 \hat Z_2 \hat C \hat \rho \hat C^{\dag}] = \Tr [\hat C^{\dag} \hat Z_1 \hat Z_2 \hat C \hat \rho ] = b.
\end{eqnarray}
Then we can asign the Pauli coefficient $a$ to all four field terms. A
representative valid Clifford operation is
\begin{eqnarray}
    \hat A_1=\hat X_2,\quad
    \hat A_2=\hat X_1,\quad
    \hat B_1=\hat Y_2,\quad
    \hat B_2=\hat Y_1.
\end{eqnarray}
This gives \(\hat B_1\hat B_2=\hat Y_1\hat Y_2\), and hence
\begin{align}
    \min_{C}\left\{ E\left( C \hat{\rho} C^\dag\right)\right\}
    =-b-2a(\abs{g}+\abs{h})= -\frac12 - \sqrt2(\abs{g}+\abs{h}).
\end{align}

On the other branch, we asign 
\begin{eqnarray}
    \Tr [\hat Z_1 \hat Z_2 \hat C \hat \rho \hat C^{\dag}] = \Tr [\hat C^{\dag} \hat Z_1 \hat Z_2 \hat C \hat \rho ] = a.
\end{eqnarray}
Without lost of generality, let us consider \(\hat B_1\hat B_2=\hat X_1\) from the $a$-weight Pauli strings $(X_1,Y_1,X_2,Y_2)$. Then both $\hat A_1$ and $\hat A_2$ must anticommute with $\hat X_1$. Among the $a$-weight Pauli strings $(X_1,Y_1,X_2,Y_2)$, only $\hat Y_1$ anticommutes with $\hat X_1$. Thus at most one of
$\hat A_1,\hat A_2$ can have weight $a$, and the other can have weight
at most $b$. Therefore,
\begin{align}
    \Tr [(\hat X_1 + \hat X_2) \hat C \hat \rho \hat C^{\dag}] = \Tr [\hat C^{\dag} (\hat X_1 + \hat X_2) \hat C \hat \rho ] \le a+b.
\end{align}
Similarly, since $\hat B_1\hat B_2=\hat X_1$, the two commuting Pauli
strings $\hat B_1,\hat B_2$ cannot have the same weight $a$, and hence
\begin{align}
    \Tr [(\hat Z_1 + \hat Z_2) \hat C \hat \rho \hat C^{\dag}] = \Tr [\hat C^{\dag} (\hat Z_1 + \hat Z_2) \hat C \hat \rho ] \le a+b.
\end{align}
These bounds are saturated, for example, by the valid Clifford transformation
\begin{align}
    \hat A_1=\hat Y_1\hat X_2,\quad
    \hat A_2=\hat Y_1,\quad
    \hat B_1=\hat Y_2,\quad
    \hat B_2=\hat X_1\hat Y_2.
\end{align}
Indeed, $\hat B_1\hat B_2=\hat X_1$, and we have
\begin{align}
    \Tr [\hat C^{\dag} (\hat X_1 + \hat X_2) \hat C \hat \rho ] =\Tr [\hat C^{\dag} (\hat Z_1 + \hat Z_2) \hat C \hat \rho ] = a+b.
\end{align}
Therefore the Clifford minimized energy in this branch is given by
\begin{align}
    \min_{C}\left\{ E\left( C \hat{\rho} C^\dag\right)\right\}
    = -a - (a+b)(\abs{g} + \abs{h}).
\end{align}

Since the initial energy is $\sqrt{2}g$, the Clifford ergotropy is given by
\begin{eqnarray}
    \mathcal E_{\rm Cl}(\hat\rho)
    =
    \sqrt2\,g+\mathcal B(g,h),
\end{eqnarray}
where
\begin{eqnarray}
    \mathcal B(g,h)
    =
    \begin{cases}
    \dfrac{1}{\sqrt2}
    +
    \left(\dfrac{1}{\sqrt2}+\dfrac12\right)s,
    & s\le 1,\\[3mm]
    \dfrac12+\sqrt2\,s,
    & s\ge 1,
    \end{cases}
\end{eqnarray}
with $s= \abs g + \abs h$.

A crossing between the two branches occurs at $\abs g + \abs h =1$. This explains the cusp at $|g|=1-|h|$ in the Clifford-optimized
landscape. In addition, the absolute values $\abs g$ and $\abs h $ generate cusps at $g=0$ and $h=0$, respectively.

\section{Proof of the extensive infinite-order filtered SRE for typical states}
Here, we show that typical states have  an extensive infinite-order filtered SRE $M_\infty$.
For this purpose, we evaluate
\begin{align}
\mathrm{Prob}\left(r_1\geq  \epsilon\right)=\mathrm{Prob}\left(\max_\mu|\rho_{\mu}|\geq  \epsilon\right),
\end{align}
from above,
where $\mathrm{Prob}$ is the measure for the Haar-random states drawn from the entire Hilbert space.
Due to the measure concentration, we first  have~\cite{popescu2006entanglement,reimann2015,hamazaki2018,meckes2019random}
\begin{align}
\mathrm{Prob}\left(\max_\mu|\rho_{\mu}|\geq  \epsilon\right)
&\leq d^2\mathrm{Prob}\left(|\rho_{\mu}|\geq  \epsilon\right)\nonumber\\
&\leq d^2 e^\pi\exp\left(-\frac{d\epsilon^2}{8}\right),
\end{align}
where we note that the Haar average of $\rho_\mu$ is zero.
Note that the coefficients are slightly different from Ref.~\cite{popescu2006entanglement,reimann2015,hamazaki2018}, because we adopt the measure-concentration result in \cite{meckes2019random}, but the difference is not important for our purpose.

Now, we take $\epsilon=\sqrt{16a\ln d/d}$, where $a$ is an arbitrary real number satisfying $a>1$.
Then, we have
\begin{align}
\mathrm{Prob}\left(r_1\geq  \sqrt{\frac{16a\ln d}{d}}\right)\leq \frac{e^\pi}{d^{2(a-1)}}.
\end{align}
This means that, for large $N$, the probability that $r_1$ is not exponentially small with $N$ (up to the polynomial correction), especially as $\sim 1/\sqrt{d}$, is exponentially small. 
In other words, typical states have  the extensive infinite-order filtered SRE $M_\infty=-2\ln r_1\simeq N\ln 2$.

\section{Proof of $E(C\hat{\rho}_\mathrm{typ}^{E_0}C^\dag)\simeq E(C\hat{\rho}_\mathrm{mic}^{E_0}C^\dag)$}
Here, we will evaluate
\begin{align}
\mathrm{Prob}_{E_0}\left(|E(C\hat{\rho}_\mathrm{typ}^{E_0}C^\dag)- E(C\hat{\rho}_\mathrm{mic}^{E_0}C^\dag)|\geq  \epsilon\right)
\end{align}
from above, where $\mathrm{Prob}_{E_0}$ is the measure for the Haar-random states drawn from the microcanonical energy shell at energy $E_0$, whose dimension is denoted as $d_\mathrm{mc}=e^{\mathrm{O}(N)}$.
We first note 
\begin{align}
|E(C\hat{\rho}_\mathrm{typ}^{E_0}C^\dag)- E(C\hat{\rho}_\mathrm{mic}^{E_0}C^\dag)|
&=\left|\sum_{\ell=1}^K \eta_\ell^C H_\ell(\rho_{c(\ell)}^\mathrm{typ}-\rho_{c(\ell)}^\mathrm{mic})\right|\nonumber\\
&\leq \|\mathbf{H}\|_1\max_\mu|\rho_{\mu}^\mathrm{typ}-\rho_{\mu}^\mathrm{mic}|,
\end{align}
where $\rho_{\mu}^\mathrm{typ/mic}=\mathrm{Tr}[\hat{\rho}_\mathrm{typ/mic}^{E_0}\hat{P}_\mu]$. Note that the Haar average of $\rho_{\mu}^\mathrm{typ}$ becomes $\rho_{\mu}^\mathrm{mic}$.

Now, due to the measure concentration, we have~\cite{popescu2006entanglement,reimann2015,hamazaki2018,meckes2019random}
\begin{align}
\mathrm{Prob}_{E_0}\left(\max_\mu|\rho_{\mu}^\mathrm{typ}-\rho_{\mu}^\mathrm{mic}|\geq  \delta\right)
&\leq 4^N\mathrm{Prob}_{E_0}\left(|\rho_{\mu}^\mathrm{typ}-\rho_{\mu}^\mathrm{mic}|\geq  \delta\right)\nonumber\\
&\leq 4^N e^\pi\exp\left(-\frac{d_\mathrm{mc}\delta^2}{8}\right).
\end{align}

Combining the above, we obtain
\begin{align}
\mathrm{Prob}_{E_0}\left(|E(C\hat{\rho}_\mathrm{typ}^{E_0}C^\dag)- E(C\hat{\rho}_\mathrm{mic}^{E_0}C^\dag)|\geq  \epsilon\right)
\leq 4^N e^\pi\exp\left(-\frac{d_\mathrm{mc}\epsilon^2}{8\|\mathbf{H}\|_1^2}\right)
\end{align}
by setting $\delta =\epsilon/ \|\mathbf{H}\|_1$.
By choosing $\epsilon=d_\mathrm{mc}^{-1/3}$, we find
\begin{align}
\mathrm{Prob}_{E_0}\left(|E(C\hat{\rho}_\mathrm{typ}^{E_0}C^\dag)- E(C\hat{\rho}_\mathrm{mic}^{E_0}C^\dag)|\geq  e^{-\mathrm{O}(N)}\right)
\lesssim e^{-e^{\mathrm{O}(N)}},
\end{align}
i.e., the probability that 
$E(C\hat{\rho}_\mathrm{typ}^{E_0}C^\dag)$ and $E(C\hat{\rho}_\mathrm{mic}^{E_0}C^\dag)$ are not exponentially close is suppressed doubly exponentially.

\end{document}